\documentclass[letter,traditabstract]{aa}

\usepackage{graphicx}
\usepackage{txfonts}
\usepackage{natbib}
\bibpunct{(}{)}{;}{a}{}{,}
\begin{document}
   \title{Captured at Millimeter Wavelengths:\\ a Flare from the
   Classical T~Tauri Star DQ Tau}

   \author{D.M. Salter \inst{1}
          \and
          M.R. Hogerheijde \inst{1}
	  \and
	  G.A. Blake \inst{2}
          }

   \institute{Leiden Observatory, Leiden University,
              P.O. Box 9513, 2300 RA Leiden, The Netherlands\\
              \email{demerese@strw.leidenuniv.nl, michiel@strw.leidenuniv.nl}
         \and
             California Institute of Technology, Division of Geological and Planetary Sciences, Mail Stop 150-21, Pasadena, CA 91125, USA \\
	     \email{gab@gps.caltech.edu}
             }

   \date{Submitted August 14, 2008; Accepted October 21, 2008}

 \abstract{For several hours on 2008 April 19 the T~Tauri spectroscopic
 binary \object{DQ Tau} was observed to brighten, reaching a maximum
 detected flux of 468\,mJy and likely making it (briefly) the
 brightest object at 3\,mm in the Taurus star-forming region. We
 present the light curve of a rarely before observed millimeter flare
 originating in the region around a pre-main-sequence star, and the
 first from a classical T~Tauri star. We discuss the
 properties and nature of the flaring behavior in the context of
 pulsed accretion flows (the current picture based largely on studies
 of this object's optically variable spectrum), as well as
 magnetospheric re-connection models (a separate theory that
 predicts millimeter flares for close binaries of high
 orbital eccentricity). We believe that the flare mechanism is
 linked to the binary orbit, and therefore periodic. DQ Tau makes a
 strong case for multi-wavelength follow-up studies, performed in
 parallel, of future flares to help determine whether
 magnetospheric and dynamical interactions in a proto-binary system
 are independent. }

 


   \maketitle
%

\section{Introduction}

Observations of pre-main-sequence (PMS) stars at millimeter
wavelengths are used to study thermal emission from gas
and dust in their circumstellar disks where planets may be
forming. At these wavelengths the disks are optically thin and the
integrated flux is related directly to the total amount of cold,
circumstellar material. But on short timescales, we show how
the dominating mechanism for millimeter emission can shift to high
energy, short-period processes. We present 3\,mm 
continuum observations of the low-mass, classical T~Tauri star
(CTTS) DQ Tau, which was observed to brighten for many hours
on 2008 April 19, reaching a maximum detected flux of
468\,mJy (compared to 17\,mJy in the days both before and
after the event).

A stellar flare at these typically quiescent wavelengths, like a solar
flare in the millimeter, is believed to be the superposition of a
gyro-synchrotron spectrum with that from synchrotron emission, the
latter peaking in the sub-millimeter to far-infrared
\citep{kaufmann1986}. The first light curve for a
strong millimeter flare observed toward a young stellar
object (YSO), \object{GMR-A} in Orion, was reported by
\citet{bower2003} and attributed to magnetic activity. In protostellar
environments, scenarios where magnetic fields can interact,
re-connect, and accelerate electrons to high energies are thought to
include: single-star stellar flares as we see in the Sun;
interactions due to changes in the magnetic field lines between a star
and its accretion disk; and, colliding magnetospheres and coronal
flares between components of a multiple-star system.

While longer-term radio variability is well documented in YSOs, and can also be the result of magnetic activity like starspots, few outbursts have been observed
\citep{feigelson1999}. DQ Tau is only the fourth flare detected toward a PMS object, and the first from a CTTS. In addition to GMR-A, our observation follows the re-occuring flares from the
weak-line T~Tauri star (WTTS) \object{V773 Tau} A, studied in
depth by \citet{massi2002,massi2006,massi2008}, and a single 1.3\,cm flare from a
second YSO in Orion named \object{ORBS} by
\citet{forbrich2008}.

In Orion, follow-up study of GMR-A confirmed a photospheric
origin of gyro-synchrotron emission around a single
magnetically-active WTTS \citep{furuya2003}, whereas observations of
the deeply embedded ORBS could only limit the emitting region to a
radius $\leq$\,2\,AU \citep{forbrich2008}. In contrast, the periodic
flaring from V773 Tau A is the first extra-solar evidence for
interacting coronal flares, attributed in this case to collisions
near periastron between extended helmet streamers
from the binary components \citep{massi2008}; a phenomenon
closely analogous to the situation for DQ Tau. With a higher
orbital eccentricity ($e$\,=\,0.556, vs. 0.3 for V773 Tau A), the DQ
Tau system geometry predicts an overlap of its two protostellar
magnetospheres at each periastron passage without the need for extended
structure \citep{mathieu1997}.

\section{The DQ Tau System}

DQ Tau ($\alpha$\,=\,04:46:53.06, $\delta$\,=\,$+$17:00:00.1)
was found by \citet{mathieu1997} to be a non-eclipsing,
double-lined spectroscopic binary, comprised of two relatively
equal-mass stars ($\sim$0.65\,M$_{\odot}$) with spectral types in the
range of K7 to M1 and a robust orbital period of 15.804 days. At
closest approach, the binary separation is 8\,R$_{*}$
(=\,13\,R$_{\odot}$), roughly  four times smaller than the 30\,R$_{*}$
(=\,60\,R$_{\odot}$) separation for V773 Tau A, and less than twice
the expected T~Tauri stellar  magnetosphere size of $\sim$\,5\,R$_{*}$
\citep{shu1994,hartmann1994}. As a result of the binary
geometry and magnetospheric models, re-connection events have been 
predicted for this source \citep{mathieu1997}, but never
tested observationally. We provide evidence that this effect is
occurring and that it may be repeatable and relevant to many
similar T~Tauri systems.

The spectral energy distribution (SED) of DQ Tau is rather
typical for a CTTS. It is best fit by
\citet{mathieu1997} using a large circumbinary disk of about
0.002$-$0.02\,M$_{\odot}$. They find an additional
5\,$\times$\,10$^{-10}$\,M$_{\odot}$ of warm ($\sim$1000\,K)
circumstellar material close to the stars is
necessary to explain the near-infrared excess, but not enough to
maintain the observed accretion rate of
5\,$\times$\,10$^{-8}$\,M$_{\odot}$\,yr$^{-1}$
\citep[see][]{hartigan1995}. DQ Tau must be transporting material into
its inner regions. Both a broad bright 10\,$\mu$m silicate feature
\citep[see][]{furlan2006}, and the absence of strongly blocked
forbidden line transitions in the receding stellar flow, support
optically thin material in a region up to 0.4\,AU around the binary
\citep{basri1997,najita2003}. 

At optical wavelengths the system is known to brighten by 0.5
magnitudes coinciding with periastron, either shortly before
or at closest approach. The mechanism regulating the
optical activity is variable, as brightenings have been
observed for only 65\% of periastron encounters, and in isolated
events, multiple brightenings were seen in a single orbit
\citep{mathieu1997}. All brightenings were recorded within
an orbital phase window of 0.30 (or 5 days) centered around a phase
($\Phi$) of 0.9, where periastron defines $\Phi$\,=\,0.0, 1.0
\citep{mathieu1997}. This evidence for accretion presents an additional
puzzle, indeed for most binaries, since a large inner gap
is expected to be cleared by the resonant and tidal forces generated
by stellar movements, thereby quickly isolating a
circumbinary disk from one or two circumstellar disks. The small
binary separation of DQ Tau, however, reduces the likelihood of stable
circumstellar disks. These effects combined should speed up disk
dissipation and halt accretion processes. Instead, it is
unclear how easily material can cross the predicted gap and why some
binaries show inner cleared regions and others, like DQ Tau, do
not \citep{jensen2007}.
 
In this context, DQ Tau was the first system to be studied in
terms of pulsed accretion flows \citep{mathieu1997}. Extensive
observations of the system's optical variability and photospheric line
properties show increased veiling effects, spectral bluenings,
emergence of outflow signatures in spectral lines, and the double-line
structure in Ca\,II suggesting that bright blue-shifted material is
being emitted near the surface of the star during an event
\citep{basri1997}. All these effects at periastron are consistent with
pulsed accretion flows based on the theoretical models of
\citet{artymowicz1996} that describe how the eccentric binary orbit
periodically perturbs the outer disk, causing material to collect at
stable points near the inner rim, and then to stream across the
gap. However, the theory does not require the peak accretion
rate to occur near periastron. Instead it has been postulated
that the accretion mechanism is further regulated by strongly variable
magnetic fields \citep{mathieu1997}.


\section{CARMA Observations}

We observed DQ Tau at 115\,GHz ($\approx$\,2.7\,mm) with the Combined
Array for Research in Millimeter-wave Astronomy (CARMA) during 5
tracks in April 2008, totaling 4 hours on-source. The millimeter
array, located in eastern California (USA), provides 150 baselines of
30-350\,m in its C configuration and records one linear
polarization. Our program cycled through 5 Taurus sources during each
track, scanning each object for 12 minutes and obtaining 1-6 scans per
source, depending on the track length. A 5-minute scan of the gain
calibrator 3C111 was included every 12-24 minutes. Multiple observing
scripts allowed rotation of the source order in each track so that we
could achieve more efficient ($u,v$)-coverage per source. In addition
to the phase calibration using 3C111, we observed the planet Uranus
for flux calibration at the start of each track, as well as the radio
source 3C84 for passband calibration.

All data were processed using the MIRIAD data reduction software
program, optimized for CARMA. We calibrated each track separately
before combining the data into a final ($u,v$)-averaged continuum
image. In the inversion step a natural weighting for
the ($u,v$)-coverage was applied, and cleaning was performed down to a
cutoff of 2$\sigma$. The continuum maps are a composite of
two 500\,MHz bands covering the frequency ranges 110.94-111.41\,GHz
and 114.80-115.27\,GHz, corresponding to the lower and upper
sidebands, respectively.


  \begin{figure*}
  \centering
  \includegraphics[width=\textwidth]{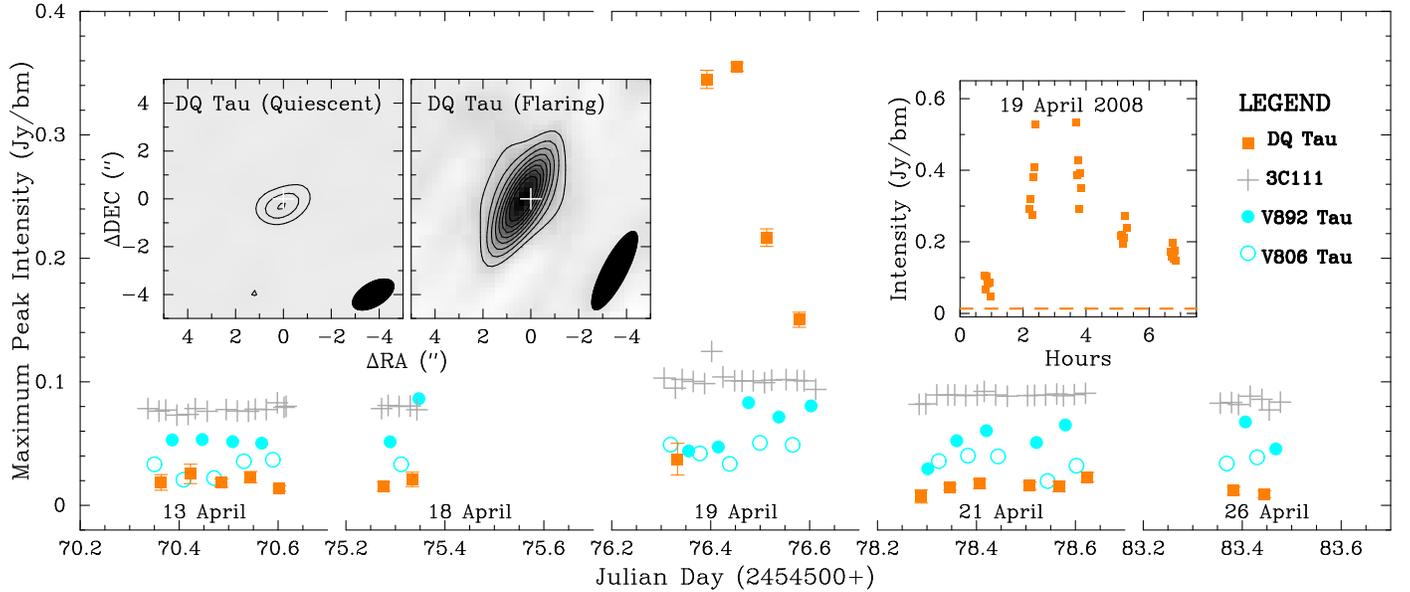}
  \caption{A plot of the maximum intensity of each scan as a function
  of Julian Day. Filled squares represent DQ Tau, circles the 2 brightest Taurus
  sources in our tracks, and crosses are the calibrator (divided by a
  factor 50). Error bars for DQ Tau are the sky rms for each scan.
  \textit{Upper Left Inset}: Continuum images of DQ Tau at
  2.7\,mm (115\,GHz). White crosses indicate the 2MASS infrared position. Pixels
  are 0.2$''$\,x\,0.2$''$ and the same grayscale is used for both
  images (ranging -3.14\,$\times$\,10$^{-2}$ to
  3.55\,$\times$\,10$^{-1}$\,Jy\,beam$^{-1}$). On the left, the
  combined image from the 4 quiescent tracks. Contours are drawn at
  3$\sigma$ levels where $\sigma$\,=\,1.45\,$\times$\,10$^{-3}$. At
  right, DQ Tau brightens by a factor of 27 on 2008 April
  19 (JD\,=\,2454576.4534), peaking at 355\,mJy\,beam$^{-1}$. We
  determine a flux of 468\,mJy for this scan.
  \textit{Upper Right Inset}: The intensity
  (Jy\,beam$^{-1}$) of the peak pixel at the location of DQ Tau
  throughout the track on 19 April, now sub-divided into 2-minute
  intervals. The horizontal dashed line indicates the maximum
  intensity found for the quiescent image. The second and third scans
  show a sharply increasing and decreasing intensity, respectively. 
  }
  \label{bigfig}
  \end{figure*}


\begin{table}
\begin{minipage}[t]{\columnwidth}
\caption{DQ Tau Flux and Intensity Data at 2.7\,mm (115\,GHz)}
\label{datatable}
\centering
\renewcommand{\footnoterule}{}  
\begin{tabular}{c c c}
\hline \hline
JD\,(2454500+)\footnote{Indicates the start of the 12-minute
scan.} & Intensity\,(mJy\,beam$^{-1}$) & Flux\,(mJy)\footnote{Determined by
a gaussian fit to the continuum image.} \\ 
\hline
   76.3330  &  37.4\,$\pm$\,15.0\footnote{Data point suffers from
   severe de-correlation at longer baselines.}  & $-$  \\ 
   76.3923  &  345.0\,$\pm$\,19.8 & 419    \\
   76.4534  &  355.1\,$\pm$\,11.4 & 468      \\
   76.5132  &  216.8\,$\pm$\,9.2  & 284     \\
   76.5789  &  150.3\,$\pm$\,6.1  & 194     \\ 
\hline                                   
   Quiescent &  13.3\,$\pm$\,1.5 & 17 \\
\hline
\end{tabular}
\end{minipage}
\end{table}


\section{Results and Analysis}

In the combined and calibrated DQ Tau image ($\sim$3 hours on-source)
we achieved a sky rms noise level of 1.45\,mJy\,beam$^{-1}$
and a beam size of 1.92$''$\,x\,0.99$''$ toward the Taurus
star-forming region, corresponding to a spatial resolution of
$\sim$140-270\,AU at a distance of 140\,pc. Since we found that
the source was considerably brighter on 19 April than during any other
observation date (Fig.~\ref{bigfig}), data for that
track were not included in this final continuum
image. And thus the quiescent values for the other nights
combined then give a peak intensity of 13.3\,mJy\,beam$^{-1}$ and a
flux, determined by a gaussian fit, of 17\,mJy.

Review of the unexpected result from 19 April
revealed that the brightening was a real effect. Many
instrumental and data processing errors could be ruled out
because none of the 4 other sources exhibited a similar effect,
the brightening persisted throughout 5 source cycles in an 8-hour
period, and, the same observing script was used for 3 of the 5
tracks. The location of the peak emission at the time of
the flare is consistent with the coordinates of the binary on all
nights and therefore no background source is
suspected. Finally, while the linear feeds do
rotate while tracking the source, we expect a flare mimicked by the
presence of strong polarization would not only be present in the
nearly identical tracks of the adjacent days, but that the emission
profile would be more symmetric about its peak.
 
To understand the nature of the DQ Tau brightening, we reduced each of
the track's five 12-minute source scans separately to produce individual
continuum maps. We then plotted the maximum source intensity
(Jy\,beam$^{-1}$ of the peak pixel) as a function of time
(Fig.~\ref{bigfig}). DQ Tau clearly exhibited a flaring behavior,
rising to a maximum brightness in $\sim$1-2 hours, coinciding with the
start of our track, and shortly thereafter, decaying (roughly
exponentially) throughout the remainder of the observation, returning
to half the observed peak intensity within a 2-hour period. 

In Fig.~\ref{bigfig} we show the results of performing the same
intensity analysis on the individual scans for three additional
sources (our two brightest Taurus sources plus the
gain calibrator). These sources appear constant in flux for all
observations (distributed over 13 days). The same is not true for DQ
Tau, which remains unresolved throughout its brightening event. At 
peak brightness its flux during our third scan on 19
April (JD\,=\,2454576.4534) was 468\,mJy, roughly 27 times 
the quiescent value (Table \ref{datatable}), likely making DQ Tau the
brightest object at 3\,mm in Taurus at the time of our observation if
we consider that the class-defining source T Tau is just
56\,$\pm$\,10\,mJy at 3\,mm \citep{ohashi1996} and the bright,
extended source HL Tau is 100\,mJy
\citep{beckwith1986}. Finally, the flare is in stark
contrast to (single-dish) measurements of DQ Tau that find 125\,mJy at
1\,mm and 91\,mJy at 1.3\,mm \citep{beckwith1991,beckwith1990}. Our
quiescent value of 17\,mJy, on the other hand, is a better fit to the
expected millimeter spectral slope. 
 
The true shape of the flare onset and its exact peak are
less precise. Larger error bars for this date are a result
of significant flagging due to strong decorrelation on longer
baselines, particularly during the first half of the
track. The gain calibrator, which has an average flux of
3.7\,Jy for each track, has a larger standard deviation
($\sigma$) for the maximum intensity of each 5-minute calibrator scan
on the flare date (0.257\,Jy\,bm$^{-1}$) than, for example, the
longest tracks on 13 April (0.095\,Jy\,bm$^{-1}$) and 21 April
(0.103\,Jy\,bm$^{-1}$). These effects do not compromise the
validity of the flare detection. For comparison, the two
brightest Taurus sources observed have a $\sigma$ of
0.013\,Jy\,bm$^{-1}$ and 0.006\,Jy\,bm$^{-1}$ for their five
12-minute scans during the flare period. 

Our flare occurs within the orbital phase
window of the known optical brightenings, although it is unknown
whether DQ Tau flared simultaneously in the optical. Using a
JD$_{0}$ value of 2449582.54 $\pm$ 0.05 and a period of 15.8043 $\pm$
0.0024 days, as published in Table 2 of \citet{mathieu1997}, we
calculate the phase of the flare peak to be 0.98, or
$\sim$7.6 hours before closest approach. If we consider the error in
the orbital period, propagated over 11 years, the flare in fact could
have peaked anywhere between 25.4 hours before periastron
($\Phi$\,=\,0.93) to 11.0 hours after ($\Phi$\,=\,0.03), 
still well within the window of optical variability. Our
other observations began at orbital phases of $-0.40$,
$-0.09$, 0.11, and 0.42 with respect to the same periastron encounter
(corresponding to $-6.32$, $-1.42$, 1.74, and 6.64 days to closest
approach), but only the track nearest periastron
($\Phi$\,=\,$-0.02$ on 19 April) exhibits a brightening
event. Therefore, we strongly suspect that the millimeter flare is
connected to the binary motions, as well as to the
optical variability, suggesting that this event is likely to be
repeatable, and even periodic.

\section{Discussion}

Of the three previous examples of flaring activity, DQ Tau
draws the strongest parallel to V773 Tau A both in system
geometry and flare properties. The similarities include: binarity
with a fairly equal mass distribution, high orbital eccentricity,
and now an observed 3\,mm flare that is nearly identical in
rise time ($\sim$2\,hours), decay time
($\sim$5\,hours, if we extrapolate in the case of DQ Tau),
peak intensity ($\sim$400\,mJy), incidence at periastron, and
ratio ($\sim$25) of maximum intensity reached with respect to
quiescent levels \citep[see][Fig.~2]{massi2006}. The most
significant difference is the lack of ongoing accretion around V773
Tau A as determined by its H$\alpha$ equivalent width.  

In the other examples, the coronal flare observed toward GMR-A
produced millimeter activity for 13 days following its intial
outburst. And while it achieved a similar peak flux
after distance correction
\citep[see][Fig.~2]{furuya2003}, the outburst activity does not appear
to be periodic (like in the case of V773 Tau A) since only a
single radio flare was detected in the course of a 7-month monitoring
campaign \citep{felli1993}. As for ORBS, the radio
flare shares a similar rise time and remains active several hours
later when the track ended
\citep[see][Fig.~1]{forbrich2008}. Unfortunately, little more is known
about the deeply embedded YSO or its flare activity. Further
comparison between these scenarios and DQ Tau is
difficult, given that we are unable to constrain the emission
polarization or its extent. 

Instead, from the millimeter perspective, the activity occurs
within the window when overlapping magnetospheres from the binary
components are expected to interact and produce flares like that
observed on 19 April. And, in the case of V773 Tau A, similar
star-star magnetospheric interactions were responsible for a nearly
identical 3\,mm flare profile. However, pulsed accretion
flows in the DQ Tau system also show a periodicity coinciding with
periastron. And while magnetospheric interaction is likely to explain
the millimeter flare, the magnetospheres do not possess the necessary
energy requirements to power the optical brightenings with the
frequency and duration observed at optical wavelengths
\citep{mathieu1997}. 

DQ Tau also possesses an additional component,
absent in the V773 Tau A system, which must be
considered: a circumbinary disk. Increased accretion at periastron
inherently requires inclusion of a star-disk interaction. When stars
accrete, material is thought to be swept from the (circumbinary) disk,
along the magnetic field lines, and onto the protostars
\citep{shu1994}. These field lines can be stretched or
compressed during accretion events, changes that can also produce
flares. Furthermore, in accretion
theory, the balance between gas pressure and
magnetic pressure often delineates the inner disk boundary. But when
a stable accretion disk is disrupted (due to the dynamical
motions of a binary) and not
available to compress the star-disk field lines, radio flaring
may occur more easily, and over larger distances, as the field lines
roam more freely \citep{basri1997}. Increased outflow signatures
during optical brightenings are consistent with increased field
alignment, if magnetic fields drive the outflows, and if pulsed
accretion can organize the lines temporarily to power those outflows
\citep{basri1997}. 

Therefore the success with each orbit of a dynamically-driven
pulsed accretion mechanism could inhibit the incidence of millimeter
activity (through compression of the field lines),
resulting in an anti-correlation between optical and
millimeter flaring. Alternatively, if the re-connection events
provide a necessary pathway for accretion processes, then the
haphazard nature of magnetic fields easily could explain the
variability in the optical flaring \citep{mathieu1997} and
the occurrence of millimeter flares and optical
brightenings is correlated. Finally, the optical
brightenings and the millimeter flares may be independent, only
sharing the common periodicity of the binary orbit. 

Simultaneous monitoring of this system at both
optical and millimeter wavelengths is important to investigate
whether magnetospheric and dynamical interactions between the stars
around periastron are strongly related. Ultimately, core fragmentation during star formation may favor
multiple-star systems like DQ Tau. Binaries comprise 65\% or more of
the stellar population in the middle of the main-sequence
\citep{duquennoy1991}. And because planets have been detected
around binaries \citep{eggenberger2004}, there is an established need
to study these systems and the impact of magnetospheric interactions
and variable accretion processes on disk evolution.

\section{Conclusions}

The PMS spectroscopic binary, DQ Tau, has a highly eccentric orbit
that should allow the stellar magnetospheres of the two components to
interact at each periastron passage. We serendipitously observed
evidence for this activity in a flare at 3\,mm, the first detected
around a CTTS. The maximum intensity recorded
(355\,mJy\,beam$^{-1}$) was 27 times the quiescent value
(13.3\,mJy\,beam$^{-1}$). The observed peak occurred roughly
7.6\,($\pm$\,18.2) hours before periastron at
$\Phi$\,=\,0.98\,$\pm$\,0.05, lasted several hours, and coincided with
the average orbital phase ($\Phi$\,=\,0.9) of the
well-documented, periodic optical brightenings. Whether a simultaneous
optical flare occurred is unknown. 

The system's optical brightenings are consistent with the
theory for pulsed accretion from the circumbinary disk onto the
protostellar surfaces, regulated by the binary orbit, and perhaps
assisted by magnetospheric interactions. The role of the
magnetospheres during accretion events can now be
studied observationally using methods established by
\citet{massi2006,massi2008} in the analysis of V773 Tau A, the first
example of coronae interactions between protostars where the authors
were able to resolve the extent of the emitting region and the
emission mechanism using very long baseline interferometry and
polarization measurements. A similar observing program aimed at DQ
Tau offers a unique opportunity to study star-disk interactions and
re-connection events directly, and to test if magnetospheric
interactions are necessary to facilitate periodic accretion.

Our findings also provide a small caution for millimeter flux points
in SEDs that could contain unrecognized flare contributions if
measured only once, with the time consuming nature of observations at
these wavelengths leading to a paucity of follow-up measurements.


\begin{acknowledgements}
	We would like to thank Jin Koda and Joanna Brown for help with
	the data collection and reduction. We would also like to thank
	the referee for many valuable comments that have helped us to improve our manuscript. Support for CARMA
	construction was derived from the states of California,
	Illinois, and Maryland, the Gordon and Betty Moore Foundation,
	the Kenneth T. and Eileen L. Norris Foundation, the Associates
	of the California Institute of Technology, and the National
	Science Foundation. Ongoing CARMA development and operations
	are supported by the National Science Foundation under a
	cooperative agreement, and by the CARMA partner
	universities. The research of DMS and MRH is supported through
	a VIDI grant from the Netherlands Organization for Scientific
	Research.
\end{acknowledgements}

\bibliographystyle{aa}
\bibliography{dqtauletter}



















\end{document}